\documentclass[epsf]{article}
\newcommand{\p}{^\prime}
\input epsf
\def\iitem#1{\noindent#1\vglue-\baselineskip\vglue-\parskip
             \hangindent=\parindent\hangafter=1}

\begin{document}

\title{MOO: A Methodology for Online Optimization through
Mining the Offline Optimum}

\author{\sl Jason W.H. Lee \qquad Y.C. Tay \qquad Anthony K.H. Tung \\
\rm National University of Singapore, \\
Kent Ridge 117543, REPUBLIC OF SINGAPORE \\
\tt tay@acm.org}

\date{Department of Mathematics Research Report No. 743\\
(June 98; revised: January 99)}
\maketitle

\centerline{\sl Abstract}
Ports, warehouses and courier services have to decide online how an arriving
task is to be served in order that cost is minimized (or profit maximized).
These operators have a wealth of historical data on task assignments; 
can these data be mined for knowledge or rules that can help
the decision-making?

MOO is a novel application of data mining to online optimization.
The idea is to mine (logged) expert decisions or the offline optimum
for rules that can be used for online decisions.
It requires little knowledge about the task distribution and cost structure,
and is applicable to a wide range of problems.

This paper presents a feasibility study of the methodology 
for the well-known $k$-server problem.
Experiments with synthetic data show that optimization can be recast as
classification of the optimum decisions;
the resulting heuristic can achieve the optimum for strong request patterns,
consistently outperforms other heuristics for weak patterns,
and is robust despite changes in cost model.

\section{Introduction}

In online optimization, a stream of tasks arrives at a system for service.
Each task must be served --- before the next arrival ---
at a cost that depends on the system's state,
which may be changed by the task.
The objective is to minimize the cost of servicing the entire task stream.

The introduction of competitive analysis [ST, KMRS] 
inspired a large body of work on online optimization in the last ten years
[BoE].
This form of analysis uses a {\it competitive ratio}
to compare the online heuristic's cost to the offline optimum
(obtained with the task stream known in advance).
In other words, the objective of the online decision algorithm
is to match the offline optimum, and this often means imitating the latter.

This objective is the basis of our proposal on a new methodology
for online optimization.
Suppose there are patterns in the task arrivals
--- i.e. task generation is constrained by a distribution;
these patterns and the cost structure in turn combine to induce 
patterns in the offline optimum solution,
and the online decision algorithm can exploit these patterns to get close
to the optimum.  Hence, the idea is:

\setlength{\parindent}{35pt}

\iitem{{\bf Step 1}}
Take a task stream (the {\it training stream}) 
that was previously generated by the distribution.

\iitem{{\bf Step 2}}
Obtain the offline optimum solution 
(i.e. the sequence of decisions for servicing the tasks).

\iitem{{\bf Step 3}}
Transform the optimum solution into a database of records.

\iitem{{\bf Step 4}}
Apply data mining to this database to extract patterns.

\iitem{{\bf Step 5}}
Use the patterns to formulate online decision rules for servicing a task
stream (the {\it test stream}) generated by the same distribution.

\setlength{\parindent}{25pt}

We call this methodology for online optimization {\it MOO},
whose essential feature is mining the offline optimum (Step 4).
This feature distinguishes MOO from the vast literature in machine learning
and database mining;
it is also different from applying algorithms for online learning
to online optimization [BB],
from using data collected online to make decisions [KMMO, FM], and 
from mining database access histories for buffer management [FLTT].
MOO's strengths are:
(1)
It is a methodology that is applicable to a wide range of problems in
online optimization (e.g. taxi assignment [FRR], 
packet routing [AAFPW], web caching [Y]).
(2)
It requires minimal knowledge about the task distribution and cost structure
(and the mining in Step 4 makes no effort to discover them).
(3)
The sort of information to be mined
(classification, association, clustering, etc.)
may vary to suit the context.
(4)
The technique for mining (item-set sampling, neural networks, etc.)
can be appropriately chosen.

On the other hand, MOO's weaknesses are:
(1)
An optimum solution for the training stream must be available.
This is an issue if no tractable algorithm is known for generating
the optimum.
MOO, however, only requires the availability of the optimum and does not
assume its tractability; it thus treats the optimum solution like an oracle.
This oracle may, in fact, be human,
in which case the methodology's objective is to approximate the expert's
performance (for this, MOO is milking the oracle offline).
Incidentally, the oracle may yield the optimum solution without providing
information about the costs.
(2)
The task distribution must be stationary [KMMO],
so that the information mined with the training stream remains relevant 
for the test stream.
(3)
MOO may need a significant amount of memory to 
store the rules for making online decisions.

To demonstrate MOO, we apply it to the {\it $k$-server problem}.
We chose this problem because
it is the prototypical and most intensively studied online problem [BoE].
It is also close to a container yard management problem 
that the Port of Singapore Authority is interested in.

The decision is cast as a classification problem,
and we use Quinlan's C4.5 to mine the optimum, 
as well as for online classification.
This software [Q] was written for machine learning,
but suffices for our purpose since the data set is not large 
and both the offline mining and online classification are fast.
However, we envisage that other applications of MOO
(e.g. using techniques other than classification,
or approximating an expert through mining historical data)
may require software that are specifically equipped 
with data mining technology [A+, H+].

We present here an experimental study of how classification
can be used for the $k$-server problem.
The objectives are:
to establish the viability of the methodology;
to explore how MOO's effectiveness is influenced by the strength of patterns,
the cost structure, the stream lengths, etc.;
and
to prepare a case for access to commercial data.

As is implicit in that third objective,
our experiments use synthetic data;
this is because
a systematic exploration of MOO's effectiveness requires controlled
experiments in which various factors can be tuned individually;
whereas
real data are affected by constraints and noise (that affect optimality),
and these get in the way of a feasibility study 
that tries to build up an understanding of the methodology.
Moreover,
gaining access to commercial data is difficult without first making a case
with synthetic data.
(As far as we know, no real data for the $k$-server problem is available
in the research community.)

The work reported here is significant in the following ways:
(1)
The experiments on synthetic data show that the methodology is feasible
--- MOO fits into the gap between the offline optimum and other online
heuristics, can come close to the optimum for strong patterns,
does well for weak patterns, and is robust with respect to the cost structure.
(2)
It shows that optimization can be recast as classification.
(3)
MOO is a novel application of a concept in data engineering to a problem
in algorithm theory,
thus serving as a bridge between the two:
This application poses challenging new problems in the analysis
of online optimization (see Section 5.2);
conversely, data mining (being an art --- consider Steps 3 to 5) 
will benefit from the algorithm community's insight
into what information to look for and how to do the mining.
(For example, the optimum solution for buffer replacement [MS]
suggests that
association rules $S\rightarrow P$ between a set of pages $S$ and a page
reference $P$ should be annotated by a ``distance'' $d$ between $S$ and $P$
mined from the reference stream, and $d$ used for buffer management [TTL].)
By offering a database perspective on online optimization,
MOO has the potential of facilitating a mutually enriching interaction
among database management, machine learning and algorithm analysis.

We first describe the $k$-server problem in Section 2.
The experimental setup is presented in Section 3
and the results examined in Section 4.
Section 5 then concludes with a summary of our observations
and poses some interesting and hard problems for this new application of 
data mining.

\section{The $k$-server problem}

The $k$-server problem is defined on a set of points with a distance
function $d$.
Conceptually, the set may be infinite but, for our experiments,
it consists of $n$ {\it nodes}.
Unlike most papers on $k$-servers, 
we do not require that $d$ satisfy the triangular inequality, 
nor that it be symmetric. 
We also do not assume that $d$ is known to the online decision algorithm.

There are $k$ {\it servers} who are positioned at different nodes.
(Some authors allow multiple servers at one node [KP].)
A {\it task} is a request that specifies a node $i$,
and is served at 0 cost if there is already a server at $i$,
or by moving a server from some node $j$ to $i$, at cost $d(j,i)$.
(Some authors allow multiple server movements per task [CL].)

A task {\it stream} is a sequence of arriving requests $T_1,\ldots,T_s$;
an {\it online solution} uses only $T_1,\ldots, T_{m-1}$ to determine
how $T_m$ is served,
while an {\it offline solution} uses $T_1,\ldots,T_s$ to determine how
each request is served.
A {\it configuration} is a set of $k$ nodes that specifies the location
of the servers before the arrival of a request.

Most algorithms in the literature for the $k$-server problem are for
special cases.
For example, Fiat et al's marking algorithm is for paging,
and Coppersmith et al's RWALK is for resistive metric spaces [FKLMSY, CDRS].
The work function algorithm [KP] is, in theory, applicable to any $k$-server
problem, but it is computationally intensive and (as far as we know)
implemented only for special cases.
In our experiments, we compare MOO to three algorithms.
If an arriving request is for node $i$ and there is no server at $i$, 
these algorithms respond as follows:

\setlength{\parindent}{50pt}

\iitem{Greedy:}
Choose a server at node $j$ for which $d(j,i)$ is minimum.

\iitem{Balance:}
Let $b_j=c_j+d(j,i)$ where $c_j$ is the cost incurred so far by the server
at node $j$; choose a server with minimum $b_j$ [MMS].

\iitem{Harmonic:}
Let $h_j=1/d(j,i)$ for each node $j$ with a server;
choose the server at $j$ with probability $h_j/\sum_r h_r$ [RS].

\noindent
Note that, unlike MOO, these three heuristics require knowledge of $d$.

\setlength{\parindent}{25pt}

\section{Experimental setup}

\subsection{Classification}

In {\it classification}, a decision tree is built from a set of {\it cases},
where each case is a tuple of {\it attribute} values.
Each attribute may be {\it discrete} (i.e. its values come from a finite set)
or {\it continuous} (i.e. the possible values form the real line).
Each case can be assigned a {\it class},
which may also be discrete (e.g. good, bad) or continuous (e.g. temperature).

Each leaf in the decision tree is a class,
and each internal node branches out based on the outcome of a test on an
attribute's value.
The tree is built from cases with known classification,
and a test case can then be classified by traversing the tree from root to 
leaf, along a path determined by the test outcomes.

For the $k$-server problem, the request distribution and distance function
induce patterns in the optimum decisions,
and MOO tries to extract these patterns for use in online assignment.
Specifically, we look for patterns that relate an assignment to the arriving
request and the configuration it sees.
Hence, the class specifies which node to move the server from,
and the classification is based on $n+1$ attributes in a case,
where one attribute specifies the arriving request and the other $n$ attributes
specify whether a node has a server;
the class and attributes are considered discrete.

(A possible alternative is to name the $k$ servers,
have the class specify the server, and
use $k$ attributes to specify the location of the servers.
With this $(k+1)$-tuple formulation of a case, however,
the classifier considers 
``server $A$ at node 1 and server $B$ at node 2''
to be different from 
``server $A$ at node 2 and server $B$ at node 1''.
This differentiation of servers is not appropriate for the $k$-server problem,
unless the cost model is changed to, say, let servers charge different costs
for movement.
It is also not appropriate to declare the class and attributes as continuous,
unless we are considering nodes on a line with a linear distance function.)

In our application of MOO, Step 2 uses network flow to solve for the 
offline optimum [CKPV];
in Step 3, this optimum is scanned to produce a file of cases,
one for each request;
Step 4 then uses C4.5 to build a decision tree with these training cases.
For a test stream, this tree is used to classify each arriving request.
This classification may be invalid, in that the tree may decide
to move a server from a node that has no server;
in this case, the server at $j$ with minimum $d(j,i)$ is chosen,
i.e. use a greedy strategy.
(If $d$ is unknown, MOO can choose a random server, say.)

\subsection{Distance function}

We choose the distance functions to test MOO's applicability for different
neighborhood structures and distance properties.
We start with $1,2,\ldots,n$ as nodes
and $d(x,x^\prime)$ given by $|x-x^\prime|$, $(x-x^\prime)^2$
and $|x-x^\prime|x^\prime$ --- 
only $|x-x\p|$ satisfies the triangular inequality,
and $|x-x\p|x\p$ is not symmetric.
We also consider $n$ nodes on a square grid with integer coordinates,
with $d((x,y),(x^\prime,y^\prime))$ given by 
$|x-x^\prime|+|y-y^\prime|$ and
$|x-x^\prime|x^\prime+|y-y^\prime|y^\prime$.

\subsection{Request generation}

The training and test streams are generated with transition matrices
in which an entry $p_{ij}$ is the probability that a request is for node $j$
given that the previous request was for node $i$.
The fraction of nonzero entries is 10--20\% for a {\it sparse} matrix and
80--90\% for a {\it dense} matrix. 
We use these matrices to generate a stream in two ways:

\setlength{\parindent}{10pt}

\iitem{$\bullet$}
A {\it 1-matrix} stream is generated with a single matrix.
This is similar to Karlin et al's markov paging,
or a random walk on Borodin et al's access graph [KPR, BIRS].

\iitem{$\bullet$}
A {\it 2-matrix} stream is generated alternately with two matrices:
$L$ requests are generated with one matrix, 
followed by $L$ requests from the other matrix; 
at the switchover, if the last request from one matrix is $i$,
then $p_{ij}$ from the other matrix is used to generate the next request.
This gives a nonhomogeneous markov chain that is a random walk on two graphs, 
in contrast to the simultaneous walks used by Fiat et al [FK, FM].
In this paper, we arbitrarily fix $L$ to be 10.
The purpose of using a 2-matrix stream is to see how MOO reacts to a mixture
of request patterns.

\setlength{\parindent}{25pt}

\noindent
An example of a matrix and a stream that it generates 
are given in the Appendix.

\begin{figure}[tbp]

\noindent
$k=5$ servers, $n=9$ nodes on a line, 
distance function $(x-x^\prime)^2$ \hfill\break
1-matrix (sparse) stream, training length 2000, test length 2000

\vbox{
\def\tablerule{\noalign{\hrule}}
\halign 
{&\vrule#& \strut\ #
 &\vrule#& \ #\  
 &\vrule#& \ #\  
 &\vrule#& \ #\  
 &\vrule#& \ #\  
 &\vrule#& \ #\  
 &\vrule#& \ #\  
 &\vrule#\cr\tablerule
& && optimum	&& \multispan7 $\underline{\rm \phantom{XXXXXXXXXXXX} 
		competitive\ ratio \phantom{XXXXXXXXXXX}}$ && invalid &\cr
& && cost  && MOO  && Greedy && Balance && Harmonic && assignment &\cr\tablerule
& $S_1$ && 402/408/\underbar{381}   && 1.00/1.01/\underbar{1.00}
	&& 1.36/1.75/\underbar{2.13} && 2.29/2.31/\underbar{2.30}
	&& 5.58/5.22/\underbar{4.93} && 0/0/\underbar{0} &\cr\tablerule
& $S_2$ && 90/113/104  && 1.09/3.04/1.30	&& 4.93/4.76/1.62 
	&& 3.00/1.98/1.92	&& 5.21/5.83/5.40 && 13/101/2 &\cr
}
\hrule}
\noindent
$S_1$ and $S_2$ are different matrices.
A triple $x/y/z$ for row $S_i$ gives the results from three task streams 
generated with $S_i$. 
For MOO, the first stream is used as the training stream, 
and all three are used as test streams; 
$x$ is the result for the training stream used as test stream
(this is why we have the same length for training and test streams).\hfill\break
The underlined numbers are results for one run (i.e. one task stream) of $S_1$.
The competitive ratio is cost incurred by an algorithm for a run divided
by the optimum cost for that run.
The last column reports the number of times the MOO classifier makes an
invalid server assignment.

\vglue 5pt
\centerline{{\bf Table 1}\quad  
		For strong patterns, MOO can be close to the optimum.}
\vglue 5pt
\hrule
\vglue -5 pt
\end{figure}

\section{Experimental results}

There are several variables in our experimental setup:
$k$, $n$, line/grid, distance, sparse/dense, pattern mixture,
starting configuration and stream length.
The stream length $s$ is the most crucial because the
offline optimum has complexity $O(ks^2)$ ---
on a 167MHz UltraSPARC, it can take 7 minutes for $s=2000$
and 1 hour for $s=2500$.
The time complexity is compounded by the large memory required
to store the network for finding the optimum
--- we have only one machine with sufficient main memory.

If we choose $s$ large enough for the optimum and heuristics 
to all reach steady state, the time commitment would be overwhelming.
Instead, in most cases, we set $s$ just large enough
that conclusions can already be drawn, 
despite significant statistical variations for any particular solution.
(This is similar to analysis of variance in statistics,
where one can separate the means of two variables if the variation of each
is ``smaller'' than the separation.)

With the bottleneck of one workstation generating the results,
we have chosen a small number of experiments that cut through the myriad
possible combinations of variables.
We concede that the data may be insufficient to support some of our conclusions,
so these should be regarded as tentative insight
rather than authoritative conclusions.

\subsection{Nodes on a line}

Table 1 presents an experiment with a strong pattern in the stream of requests
coming to 5 servers for 9 nodes on a line, 
with a $d$ that violates triangular inequality.
After 2000 requests, the fluctuations are small enough for us to draw
some conclusions.

First, the average optimum cost per request is less than 1,
and this is because most requests are for a node that already has a server.
Second, the competitive ratios for a fixed request distribution can be 
significantly smaller than the $k$-server bound [MMS];
this is similar to previous observations [BaE, FR].
Third, MOO can achieve the optimum --- 
the sparse matrix induces a strong pattern in the offline optimum solution, 
and this pattern is captured in the decision tree used by MOO.

The starting configuration used in the three runs are the same for $S_1$,
but different for $S_2$.
The results for $S_2$ show that the configuration can have a strong effect
--- the heuristics' performance ordering and competitive ratios 
both become erratic.
In contrast, the ordering for the three runs of $S_1$ are the same,
and the ratios are reasonably stable except for Greedy,
which is sensitive to the stream instance.
To factor in the effect of the starting configuration,
this configuration is henceforth changed from run to run,
unless otherwise stated.

Despite the erratic results for $S_2$ and the fact that MOO uses a greedy
strategy whenever the classifier makes an invalid assignment,
MOO has a significantly smaller ratio that Greedy,
thus showing the contribution from data mining.
A check shows that the trees are small but unintuitive
--- an example is given in the Appendix ---
since they imitate the offline optimum (which ``sees'' future requests).

\begin{figure}[tbp]
\noindent
$k=5$ servers, $n=9$ nodes on a line, 
distance function $(x-x^\prime)^2$ \hfill\break
1-matrix (dense) stream, training length 2000, test length 2000

\vbox{
\def\tablerule{\noalign{\hrule}}
\halign 
{&\vrule#& \strut\ #  
 &\vrule#& \ #\  
 &\vrule#& \ #\  
 &\vrule#& \ #\  
 &\vrule#& \ #\  
 &\vrule#& \ #\  
 &\vrule#& \ #\  
 &\vrule#\cr\tablerule
& && optimum	&& \multispan7 $\underline{\rm \phantom{XXXXXXXXXXXX} 
		competitive\ ratio \phantom{XXXXXXXXXXX}}$ && invalid &\cr
& && cost  && MOO  && Greedy && Balance && Harmonic && assignment &\cr\tablerule
& $D_1$ && 715/687/728	&& 1.16/1.21/1.20	&& 1.28/1.27/2.09	
	&& 1.72/1.85/1.66	&& 4.24/4.40/4.26 && 1/1/0 &\cr\tablerule
& $D_2$ && 684/692/732  && 1.19/1.22/1.18	&& 1.94/1.44/1.29 
	&& 1.72/1.88/1.87	&& 3.71/4.70/4.37 && 1/10/0 &\cr
}
\hrule}

\vglue 5pt
\centerline{{\bf Table 2}\quad  For weak patterns, MOO is best.}
\vglue 5pt
\hrule
\vglue -5pt
\end{figure}

In Table 1, MOO can get close to the optimum because the patterns are strong.
For a dense matrix, the pattern is much weaker.
Nonetheless, Table 2 shows that MOO has the smallest ratio,
and the invalid assignments are surprisingly few.
Further,
the difference in starting configurations between the training and test 
streams does not have a big effect on MOO's results, 
in contrast to the results for a strong pattern
(recall: the starting configurations in Table 1
are the same for 1.00/1.01/1.00 and different for 1.09/3.04/1.30).

The number of potential cases for the classifier is $n {n\choose k}$, 
which is 1134 and comparable to the training length (2000) for Table 2.
Even so, the performance ordering and ratios are reasonably stable, 
except for Greedy;
when we tested the heuristics again with the runs using the same 
starting configuration,
fluctuation in Greedy's ratios narrowed down considerably,
thus indicating that Greedy remains sensitive to the starting configuration
for weak patterns.
The decision trees, though bigger than the two for Table 1, remain small: 
the tree for $D_1$ is 3Kbytes
and has only 27 decision nodes.

\begin{figure}[tbp]
\noindent
\vbox{\tabskip=1em 
\halign{& #\hfil\cr
\epsfbox{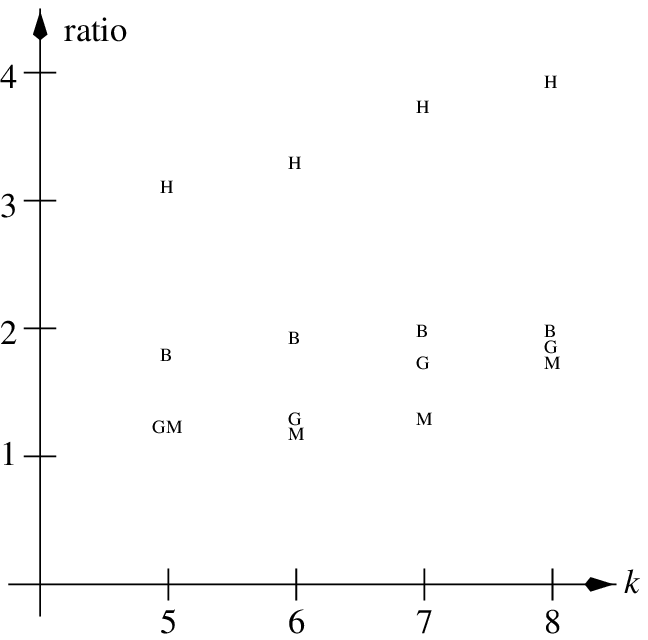} & \epsfbox{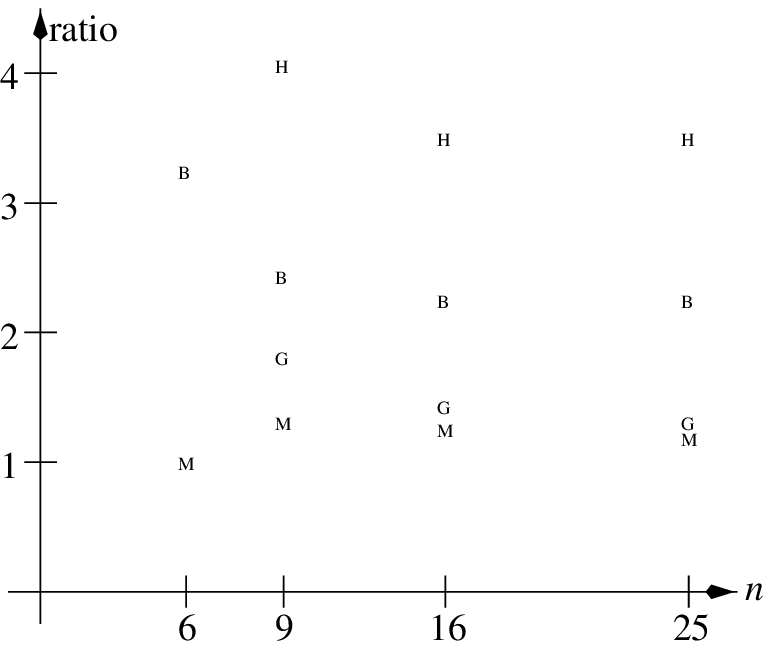} \cr
\qquad $n=9$ nodes, distance $|x-x^\prime|$ & 
\qquad $k=5$ servers, distance $|x-x^\prime|x^\prime$ \cr
\qquad 2-matrix (sparse-dense) stream &
\qquad 2-matrix (dense-dense) stream \cr
\qquad stream length 2000 &
\qquad stream length varies with $n$ \cr
\qquad H is for Harmonic, B for Balance, &
\qquad at $n=6$, H is 9.6 and G is 10.6\cr
\qquad G for Greedy, M for MOO &
\qquad \hglue 1 true cm  \cr
{\bf Figure 1}\quad MOO fits into the gap &
            {\bf Figure 2}\quad MOO stays close to optimum \cr
\hglue 1.8 true cm   between Greedy and optimum. &
\hglue 1.9 true cm   for all $n$. \cr
}}
\vskip 5pt
\hrule
\vskip -5pt
\end{figure}

All heuristics are trivially optimum if $k=1$,
but the gap between existing heuristics and the optimum should open up
as $k$ increases; 
to prove its worth, MOO must fit into this gap.

In Figure 1 (and the following graphs), 
each data point is the average of 6 runs.
It shows that, for a 2-matrix stream and distance $|x-x^\prime|$,
the gap between Greedy and optimum opens up at $k=5$ for $n=9$,
and MOO does fit into the gap.
At $k=5$ for $|x-x^\prime|$, the difference between MOO and Greedy is
negligible (if we consider the average ratio over 6 runs;
Greedy's ratio is smaller in some runs and MOO's smaller in others).
In contrast, Tables 1 and 2 show that MOO's ratios are noticeably smaller
than Greedy's at $k=5$ for $(x-x^\prime)^2$,
which penalizes large movements.
The gaps among the heuristics open further at $k=5$ and $n=9$ for $|x-x\p|x\p$
in Figure 2.

The alternation between strong and weak patterns does not affect
MOO's ability to outperform the other heuristics in Figure 1,
and Figure 2 shows this remains so for alternating between two weak patterns.
In fact, unlike Harmonic and Balance,
MOO stays close to the optimum as $n$ scales up,
thus demonstrating again its ability to learn from the optimum solution.

For an asymmetrical and punitive $|x-x\p|x\p$,
the ``right'' server placement is important to being close to optimum
for small $n$,
so Greedy's simplistic strategy does poorly there.
For large $n$, even the optimum has its servers spread out,
and the violation of the triangular inequality favors incremental
server movements,
thus making it possible for Greedy to get close to the optimum.

\begin{figure}[tbp]
\noindent
\vbox{\tabskip=1em 
\halign{& #\hfil\cr
\epsfbox{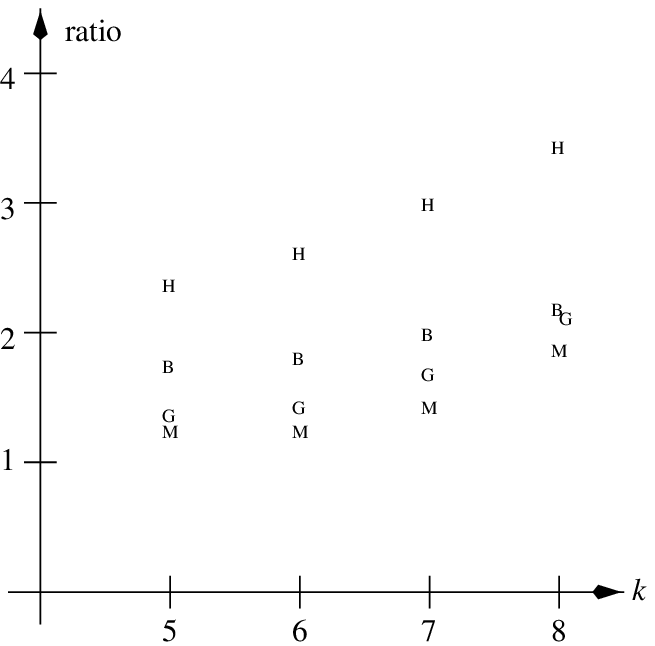} & \epsfbox{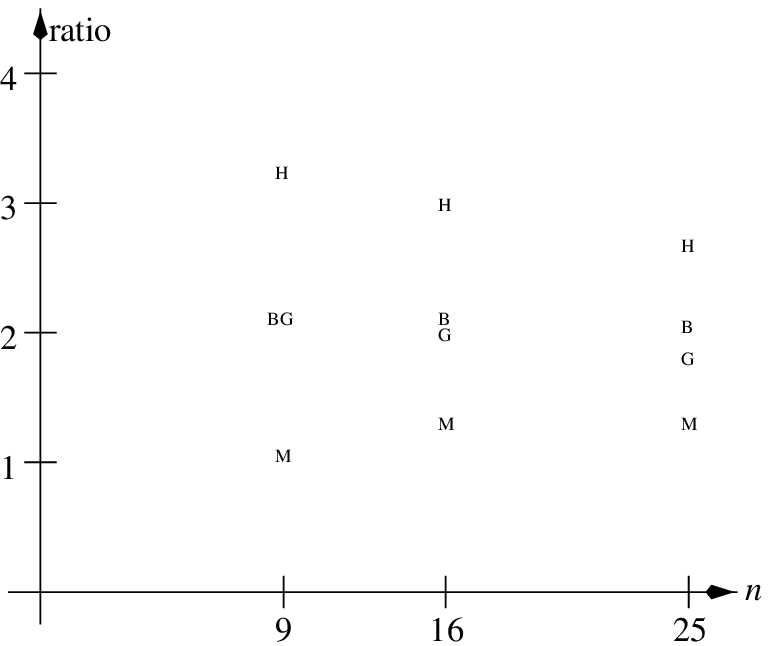} \cr
\qquad $n=9$, distance $|x-x^\prime|+|y-y\p|$ & 
\qquad $k=5$, distance $|x-x^\prime|x^\prime+|y-y\p|y\p$ \cr
\qquad same stream and starting configuration &
\qquad same stream and starting configuration \cr
\qquad as Figure 1 &
\qquad as Figure 2 \cr
{\bf Figure 3}\quad For a grid, & 
            {\bf Figure 4}\quad For a grid, \cr
\hglue 1.8 true cm   MOO still fits in the gap. &
\hglue 1.9 true cm   MOO still stays close to optimum. \cr
}}
\vskip 5pt
\hrule
\vskip -5pt
\end{figure}

\subsection{Nodes on a grid}

Intuitively, a heuristic should incur lower costs if nodes have more neighbors,
but its ratio can increase because
the optimum may make better use of the neighbors in reducing its cost.

Figure 3 shows the results of repeating the runs for Figure 1
--- same starting configurations and request streams --- 
on a grid (instead of a line).
Harmonic does perform better,
but the effect on the ratios for Balance and Greedy is mixed.
A check (of the detailed data) shows that, contrary to our intuition,
their costs are sometimes higher for the grid.
It appears that the increase in the number of neighbors
also leads Balance and Greedy to make short-sighted moves that
raise costs eventually.
In any case, MOO remains in the gap between Greedy and optimum
when $k$ increases.

Similar results hold when $n$ is varied.
Comparing Figures 2 and 4,
we see that the ratios for a grid are noticeably smaller for Harmonic
but larger for Greedy.
A check shows that costs are lower (often by an order of magnitude),
so all solutions benefit from having more neighbors 
when $d$ is $|x-x\p|x\p+|y-y\p|y\p$.
However, the spreading-out effect that allows Greedy to get close to
the optimum in Figure 2 is less for a grid,
so Greedy is further from the optimum in Figure 4.
Again, we see the gap among the heuristics opening up at $k=5$ and $n=9$
when $d$ changes from $|x-x\p|+|y-y\p|$ to $|x-x\p|x\p+|y-y\p|y\p$.

MOO, on the other hand, stays close to optimum, like in Figure 2.
The detailed data show that there are at most 2 invalid assignments 
(that are resolved greedily) at $n=9$ and less than $12\%$ such assignments
at $n=25$; hence, MOO relies mostly on the decision tree,
which has successfully captured the optimum solution
even though the requests are a mixture of two weak patterns.

\section{Conclusion}

\subsection{Summary}

We now summarize our observations:

\iitem{$\bullet$}
MOO fits into the gap between the offline optimum and other online heuristics
(Figures 1--4).
For a strong pattern, MOO can be close to optimum,
but may lose to other heuristics because of sensitivity to the starting
configuration (Table 1).
MOO does well even if the requests have a weak pattern (Table 2)
or alternate between patterns (Figures 1--4).

\iitem{$\bullet$}
MOO outperforms the other heuristics even if the distances are asymmetric
(Figures 2 and 4) or violate the triangular inequality (Tables 1 and 2).
Increasing the number of neighbors can increase costs,
but MOO's ratios remain stable (Figures 1 and 3, 2 and 4).

\iitem{$\bullet$}
MOO stays close to the optimum as $n$ varies (Figures 2 and 4).

\iitem{$\bullet$}
The classifier can get an effective decision tree even for 
relatively short stream lengths, the trees are small and the mining (Step 4)
is fast (sub-second).

\subsection{Challenging issues}

MOO poses some challenging problems for 
this new application of data mining:

\iitem{$\bullet$}
How to analyze the competitive ratios produced with data mining?

\iitem{$\bullet$}
For the $k$-server problem,
why does MOO perform well for weak patterns and short training
streams?
(For the buffer replacement problem, mining can produce good results
even if the requests are a mixture of 100 patterns [TTL].)

\iitem{$\bullet$}
What sort of data mining would be appropriate for 
web caching, video-on-demand, etc.?

\vskip25pt\noindent
{\bf Acknowledgment}

\noindent
Many thanks to C.P. Teo for his help with network flow
and Hongjun Lu for his comments.

\subsection{References}

\setlength{\parindent}{2.0 cm}

\iitem{[A+]}
R. Agrawal, M. Mehta, J. Shafer, R. Srikant, A. Arning and T. Bollinger,
{\it The Quest data mining system},
Proc. KDD, Portland, OR (Aug. 1996), 244--249.

\iitem{[AAFPW]}
J. Aspnes, Y. Azar, A. Fiat, S. Plotkin and O. Waarts,
{\it On-line load balancing with applications to machine scheduling and
virtual circuit routing},
Proc. STOC, San Diego, CA (May 1993), 623--630.

\iitem{[BB]}
A. Blum and C. Burch,
{\it On-line learning and the metrical task system problem},
Proc. COLT, Nashville, TN (July 1997), 45--53.

\iitem{[BaE]}
R. Bachrach and R. El-Yaniv,
{\it Online list accessing algorithms and their applications: recent 
empirical evidence},
Proc. SODA, New Orleans, LA (Jan. 97), 53--62.

\iitem{[BoE]}
A. Borodin and R. El-Yaniv,
{\sl Online Computation and Competitive Analysis},
Cambridge University Press, Cambridge, UK (1998).

\iitem{[BIRS]}
A. Borodin, S. Irani, P. Raghavan and B. Schieber,
{\it Competitive paging with locality of reference},
Proc. STOC, New Orleans, LA (May 1991), 249--259.

\iitem{[CDRS]}
D. Coppersmith, P. Doyle, P. Raghavan and M. Snir,
{\it Random walks on weighted graphs and applications to on-line algorithms},
J. ACM 40, 3 (July 1993), 421--453.

\iitem{[CKPV]}
M. Chrobak, H. Karloff, T. Payne and S. Vishwanathan,
{\it New results on server problems},
SIAM J. Disc. Math. 4, 2(May 1991), 172--181.

\iitem{[CL]}
M. Chrobak and L.L. Larmore,
{\it An optimal on-line algorithm for $k$-servers on trees},
SIAM J. Computing 20, 1(1991), 144--148.

\iitem{[FK]}
A. Fiat and A.R. Karlin,
{\it Randomized and multipointer paging with locality of reference},
Proc. STOC, Las Vegas, NV (May 1995), 626--634.

\iitem{[FKLMSY]}
A. Fiat, R.M. Karp, M. Luby, L.A. McGoech, D.D. Sleator and N.E. Young,
{\it Competitive paging algorithms},
J. Algorithms 12, 4(Dec. 1991), 685--699.

\iitem{[FLTT]}
L. Feng, H. Lu, Y.C. Tay and K.H. Tung,
{\it Buffer management in distributed database systems: 
A data mining approach},
Proc. EDBT, Valencia, Spain (Apr. 1998), 246--260.

\iitem{[FM]}
A. Fiat and M. Mendel,
{\it Truly online paging with locality of reference},
Proc. FOCS, Miami Beach, FL (Oct. 1997), 326--335.

\iitem{[FR]}
A. Fiat and Z. Rosen,
{\it Experimental studies of access graph based heuristics:
beating the LRU standard?},
Proc. SODA, New Orleans, LA (Jan. 1997), 63--72.

\iitem{[FRR]}
A. Fiat, Y. Rabani and Y. Ravid,
{\it Competitive $k$-server algorithms},
Proc. FOCS, St. Louis, MO (Oct. 1990), 454--463.

\iitem{[H+]}
J. Han, Y. Fu, W. Wang, J. Chiang, W. Gong, K. Koperski, D. Li, Y. Lu,
A. Rajan, N. Stefanovic, B. Xia and O.R. Zaiane,
{\it DBMiner: A system for mining knowledge in large relational databases},
Proc. KDD, Portland, OR (Aug. 1996), 250--255.

\iitem{[KMMO]}
A.R. Karlin, M.S. Manasse, L.A. McGeoch and S. Owicki,
{\it Competitive randomized algorithms for non-uniform problems},
Proc. SODA, San Francisco, CA (Jan. 1990), 301--309.

\iitem{[KMRS]}
A.R. Karlin, M.S. Manasse, L. Rudolph and D.D. Sleator,
{\it Competitive snoopy caching},
Algorithmica 3, 1(1988), 79--119.

\iitem{[KP]}
E. Koutsoupias and C. Papadimitriou,
{\it On the $k$-server conjecture},
Proc. STOC, Montreal, Canada (May 1994), 507--511.

\iitem{[KPR]}
A.R. Karlin, S.J. Phillips and P. Raghavan,
{\it Markov paging},
Proc. FOCS, Pittsburgh, PA (Oct. 1992), 208--217.

\iitem{[MMS]}
M.S. Manasse, L.A. McGeoch and D.D. Sleator,
{\it Competitive algorithms for on-line problems},
Proc. STOC, Chicago, IL (May 1988), 322--333.

\iitem{[MS]}
L.A. McGeoch and D.D. Sleator,
{\it A strongly competitive randomized paging algorithm},
Algorithmica 6, 6(1991), 816--825.

\iitem{[Q]}
J.R. Quinlan,
{\sl C4.5: Programs for Machine Learning},
Morgan Kaufman, San Mateo, CA (1993).

\iitem{[RS]}
P. Raghavan and M. Snir,
{\it Memory versus randomization in on-line algorithms},
Proc. ICALP, Stresa, Italy (July 1989), 687--703.

\iitem{[ST]}
D.D. Sleator and R.E. Tarjan,
{\it Amortized efficiency of list update and paging rules},
C. ACM 28, 2(Feb. 1985), 202--208.

\iitem{[T]}
K.H. Tung, 
{\it Parking in a Marina},
Honors Year Project Report, DISCS, National University of Singapore (1997).

\iitem{[TTL]}
K.H. Tung, Y.C. Tay and H. Lu,
{\it BROOM: Buffer replacement using online optimization by mining},
Proc. CIKM, Bethesda, MD (Nov. 1998), 185--192.

\iitem{[Y]}
N. Young,
{\it On-line file caching},
Proc. SODA, San Francisco, CA (Jan. 1998), 82--86.

\newpage
\section{Appendix}

$$\bordermatrix{
  &   0   &  1    & 2    & 3    & 4    & 5    & 6    & 7    & 8    \cr
0 &  0.00 &  0.45 & 0.00 & 0.55 & 0.00 & 0.00 & 0.00 & 0.00 & 0.00 \cr
1 &  0.00 &  0.00 & 0.00 & 0.58 & 0.00 & 0.00 & 0.00 & 0.00 & 0.42 \cr
2 &  0.31 &  0.69 & 0.00 & 0.00 & 0.00 & 0.00 & 0.00 & 0.00 & 0.00 \cr
3 &  0.00 &  0.00 & 0.00 & 0.00 & 0.00 & 1.00 & 0.00 & 0.00 & 0.00 \cr
4 &  1.00 &  0.00 & 0.00 & 0.00 & 0.00 & 0.00 & 0.00 & 0.00 & 0.00 \cr
5 &  0.00 &  0.00 & 0.00 & 0.00 & 0.00 & 0.00 & 0.00 & 0.02 & 0.98 \cr
6 &  1.00 &  0.00 & 0.00 & 0.00 & 0.00 & 0.00 & 0.00 & 0.00 & 0.00 \cr
7 &  0.00 &  0.00 & 0.35 & 0.00 & 0.62 & 0.03 & 0.00 & 0.00 & 0.00 \cr
8 &  0.00 &  0.47 & 0.00 & 0.00 & 0.00 & 0.00 & 0.53 & 0.00 & 0.00 \cr}$$

\centerline{{\bf Figure A.1}\quad Sparse matrix $S_1$ of Table 1.}

\vglue 10pt

\noindent
1 8 6 0 1 3 5 8 6 0 3 5 8 1 3 5 8 6 0 1 3 5 8 1 3 5 7 2 1 3 5 8 1 3 5 8 6 0 1 8 6 0 1 8 1 3 5 8 1 3 5 8 1 

\vglue 5pt
\centerline{{\bf Figure A.2}\quad $S_1$ generates a strong pattern.}

\vglue 10pt

\vbox{\tabskip=1em 
\halign{& #\hfil\cr
Request from = 2: 3 	  &\cr
Request from = 4: 5 	  &\cr
Request from = 7: 8 	  &\cr
Request from = 0:         &\cr
$|$\quad   Node 0 status = 0: 1  &
	{\tt // this tree has depth 1 only }\cr
$|$\quad   Node 0 status = 1: 0  &
	{\tt // weaker patterns induce deeper trees }\cr
Request from = 1:         &\cr
$|$\quad   Node 0 status = 0: 1  &\cr
$|$\quad   Node 0 status = 1: 0  &
	{\tt // how to read C4.5's decision tree:}\cr
Request from = 3:         &
	{\tt // if the request is for node 3 }\cr
$|$\quad   Node 2 status = 0: 3  &
	{\tt // then (a) if no server is at 2, then use server at 3}\cr
$|$\quad   Node 2 status = 1: 2  &
	{\tt // \qquad \ (b) else move the server from 2 }\cr
Request from = 5:         &
	{\tt // note: the tree is used only if no server }\cr
$|$\quad   Node 5 status = 0: 4  &
	{\tt //\qquad\qquad is at the requested node}\cr
$|$\quad   Node 5 status = 1: 5  &
	{\tt //\qquad\qquad so (a) is an invalid assignment }\cr
Request from = 6:         &
	{\tt //\qquad\qquad and (b) will not put two servers at 3 }\cr
$|$\quad   Node 6 status = 0: 5  &\cr
$|$\quad   Node 6 status = 1: 6  &\cr
Request from = 8:         &
	{\tt // this tree always assigns a server from a neighboring node }\cr
$|$\quad   Node 8 status = 0: 7  &
	{\tt // in agreement with $d$ in Table 1}\cr
$|$\quad   Node 8 status = 1: 8  &
	{\tt // which favors incremental movements }\cr
}}

\vglue 5pt\noindent
Note that C4.5 (appropriately) selects the request to be the root.
However, the rest of the tree is unintuitive,
since the tree is mined from an offline optimum that ``sees'' future requests.

\vglue 5pt
\centerline{{\bf Figure A.3}\quad
Decision tree from an optimum solution for a sequence generated with $S_1$.}

\end{document}